\documentclass[twoside]{article}
\usepackage{fleqn}
\usepackage{espcrc2}
\usepackage{amsmath}

\usepackage{epsfig}

\def\Order#1{${\cal O}(#1$)}

\def\Order#1{${\cal O}(#1)$}

\def\KK{${\cal KK}$}

\def\bbeta{\bar{\beta}}
\def\hbeta{\hat{\beta}}
\newcommand{\sfac}{\mathfrak{s}}
\usepackage{amssymb}
\usepackage{euscript}
\newcommand{\Meu}{\EuScript{M}}
\title{Coherent Exclusive Exponentiation for\\
       Precision Monte Carlo Calculations%
\thanks{Work supported in part by 
        the US DoE contracts DE-FG05-91ER40627 and DE-AC03-76SF00515,
        Polish Government grants
        KBN 2P03B08414, %%%<-- Marek
        KBN 2P03B14715, %%%<-- Zbyszek
        the Maria Sk\l{}odowska-Curie Joint Fund II PAA/DOE-97-316,
        and the Polish--French Collaboration within IN2P3 through LAPP Annecy.}
}

\author{S. Jadach%
        \address{Department of Physics and Astronomy,\\
                 The University of Tennessee, Knoxville, Tennessee 37996-1200, USA\\
                 and
                 Institute of Nuclear Physics, Cracow, ul. Kawiory 26A, Poland},
        B.F.L. Ward%
        \address{Department of Physics and Astronomy,\\
                 The University of Tennessee, Knoxville, Tennessee 37996-1200, USA}
        and
        Z. W\c{a}s%
        \address{Institute of Nuclear Physics, Cracow, ul. Kawiory 26A, Poland},
}

\begin{document}
%=====================================

\begin{abstract}
In this contribution we give a short overview of the
new Coherent Exclusive Exponentiation (CEEX)
which is implemented in the new \KK MC event generator
for the process $e^+e^-\to f\bar{f} +n\gamma$, $f=\mu,\tau,d,u,s,c,b$
with validity for center of mass energies from $\tau$ lepton threshold to 1TeV,
that is for LEP1, LEP2, SLC, future Linear Colliders, $b,c,\tau$-factories etc.
In CEEX effects due to photon emission from initial beams and outgoing fermions
are calculated in QED up to second-order, including all interference effects.
Electroweak corrections are included in first-order, at the amplitude level.
Beams can be polarized longitudinally and transversely,
and all spin correlations are incorporated in an exact manner.
We describe briefly the essence of CEEX as compared with older exclusive
variants of the exponentiation (EEX)
and present samples of the numerical results,
concentrating on the question of the technical and physical precision for the total cross section
and for the charge asymmetry.
New results on the interference between initial and final state photon
emission at LEP2 energies are also shown.
\end{abstract}

\maketitle

%%%%%%%%%%%%%%%%%%%%%%%%%%%%%%%%%%%%%%%%%%%%%%%%%%%%%%%%%%%%%%%%%
\section{Introduction}
%%%%%%%%%%%%%%%%%%%%%%%%%%%%%%%%%%%%%%%%%%%%%%%%%%%%%%%%%%%%%%%%%
At the end of LEP2 operation the total cross section for 
the process $e^-e^+\to f\bar{f}+n\gamma$
will have to be calculated with the precision $0.2\%-1\%$, depending on event selection.
The arbitrary differential distributions have to be calculated also
with the corresponding precision.
In future linear colliders (LC's) the precision requirement can be even more demanding.
This is especially true for high luminosity linear colliders, like in the case of TESLA.
The above new requirements necessitate development of the new calculational framework 
for the QED corrections and the construction of new dedicated MC programs.
We present here a new effort in this direction.
This report is based on refs.~\cite{ceex1:1999,gps:1998,ceex2:2000}
and the Monte Carlo program is described in ref.~\cite{kkcpc:1999}.

%%%%%%%%%%%%%%%%%%%%%%%%%%%%%%%%%%%%%%%%%%%%%%%%%%%%%%%%%%%%%%%%%%%%%%%%%%%%%
\begin{table*}[!ht]
\centering
%---------------------------
\setlength{\unitlength}{0.1mm}
\begin{picture}(1400,900)
%#####\put(0,0){\framebox( 800,800){ }}
%#####
\put( 0, 0){\makebox(0,0)[lb]{\epsfig{file=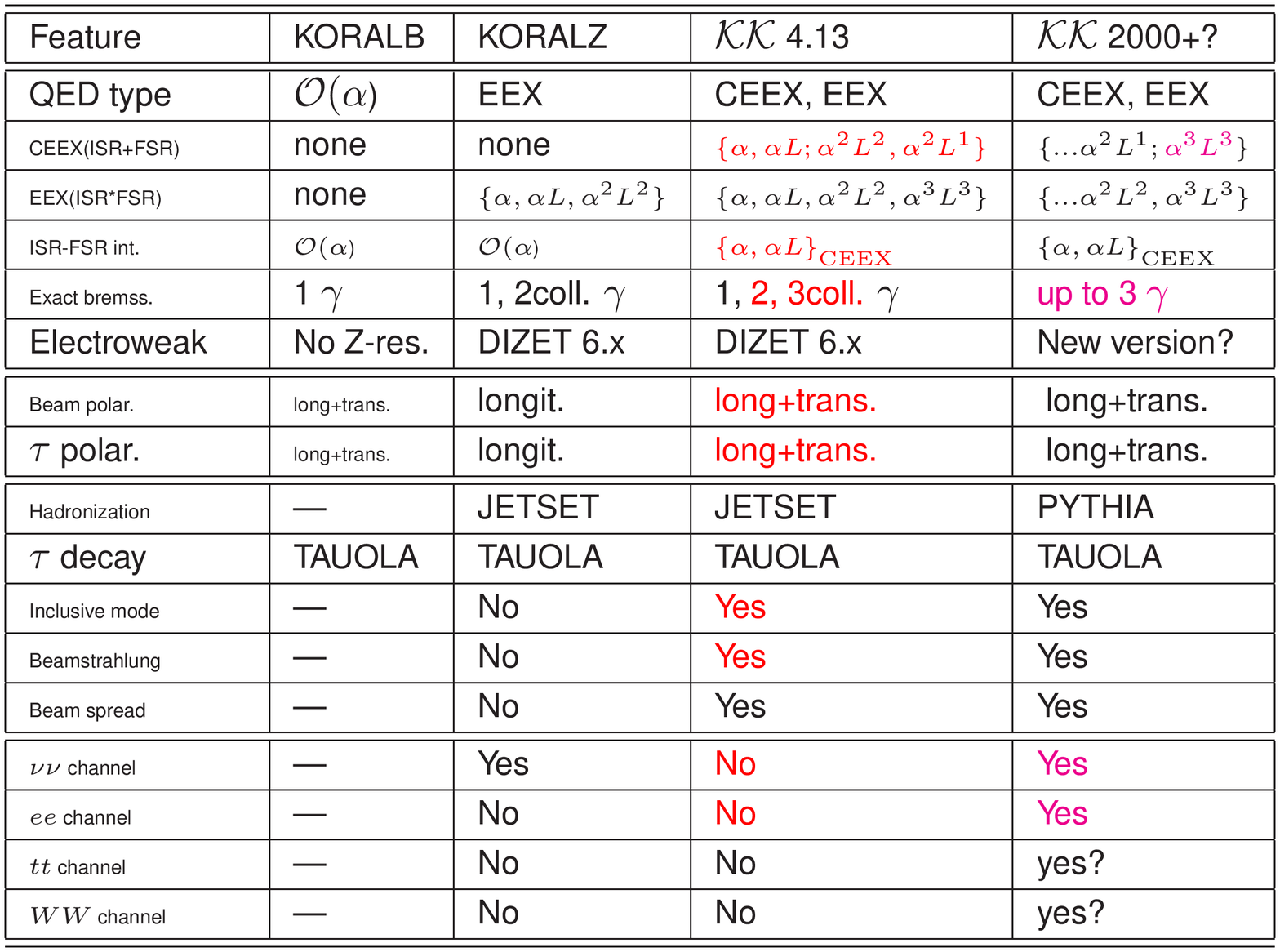,width=140mm,height=90mm}}}
\end{picture}
\caption{Overview of \KK MC event generator as compared with KORALZ and KORALB.
}
\label{tab:compare}
\end{table*}
%%%%%%%%%%%%%%%%%%%%%%%%%%%%%%%%%%%%%%%%%%%%%%%%%%%%%%%%%%%%%%%%%%%%%%%%%%%%%

%%%%%%%%%%%%%%%%%%%%%%%%%%%%%%%%%%%%%%%%%%%%%%%%%%%%%%%%%%%%%%%%%
\section{What is coherent exclusive exponentiation CEEX?}
%%%%%%%%%%%%%%%%%%%%%%%%%%%%%%%%%%%%%%%%%%%%%%%%%%%%%%%%%%%%%%%%%
The {\em exponentiation} is generally a method of summing up real and virtual photon
contributions to infinite order such that infrared (IR) divergences cancel.
The {\em exclusivity} means that the procedure of exponentiation, that is summing up
the infrared (IR) real and virtual contribution, within the standard perturbative scheme
of quantum field theory, is done at the level of the fully differential (multiphoton) cross section,
or even better, at the level of the scattering matrix element (spin amplitude),
{\em before any phase-space integration over photon momenta is done}.
The other popular type of the exponentiation is {\em inclusive} exponentiation (IEX), 
which is done at the level of inclusive distributions, structure functions, etc.
see discussion in ref.~\cite{sussex:1989}.
The classical work of  Yennie-Frautschi-Suura~\cite{yfs:1961} (YFS) represents the best
example of the exclusive exponentiation and we nickname it as EEX.
Finally, why do we use word {\em coherent?} In CEEX the
essential part of the summation of the IR real and virtual photon contributions
is done at the amplitude level.
Of course, IR cancellations occur as usual at the probability level, however,
the transition from spin amplitudes to differential cross sections,
and the phase space integration are done entirely numerically!
As a consequence of the above {\em coherent} approach it follows,
that CEEX is friendly
to coherence among Feynman diagrams, narrow resonances, interferences etc.
This is great practical advantage.
In our many previous works which led to the development of the Monte Carlo
event generators like YFS3, YFS3, KORALZ, KORALW, YFS3WW, BHLUMI, BHWIDE,
see refs.~\cite{yfs2:1990,koralz4:1994,koralw:1998,yfsww:1996,bhlumi4:1996,bhwide:1997},
we have generally employed EEX, which is closely related to the YFS work~\cite{sussex:1989}.
The CEEX is a recent development and is so far used only in 
the new \KK MC program~\cite{kkcpc:1999}.

Let us now show in a very simplified schematic way what is the the main difference
between old EEX/YFS and CEEX for the fermion pair production the process:
\begin{equation}
\begin{split}
& e^-(p_1,\lambda_1)+e^+(p_2,\lambda_2) \to \\
& f(q_1,\lambda'_1)+\bar{f}(q_2,\lambda'_2)+\gamma(k_1,\sigma_1)+...+\gamma(k_n,\sigma_n).
\end{split}
\end{equation}
The EEX total cross section is
\begin{equation}
\sigma = \sum\limits_{n=0}^\infty \;\;
          \int\limits_{m_\gamma}
          d\Phi_{n+2}\; e^{Y({m_\gamma})}
          D_n(q_1,q_2,k_1,...,k_n),
\end{equation}
where in the \Order{\alpha^1} the distributions for $n_\gamma=0,1,2$ are
\begin{equation}
\begin{split}
    &D_0          =  \bbeta_0\\
    &D_1(k_1)     =  \bbeta_0 \tilde{S}(k_1) +\bbeta_1(k_1)\\
    &D_2(k_1,k_2) =  \bbeta_0 \tilde{S}(k_1)\tilde{S}(k_2)\\
    &\qquad\qquad                +\bbeta_1(k_1)\tilde{S}(k_2)+\bbeta_1(k_2)\tilde{S}(k_1)
\end{split}
\end{equation}
and the real soft factors are defined as usual
\begin{equation}
\begin{split}
 4\pi\tilde{S}(k) &= \sum\limits_\sigma |\sfac_\sigma(k)|^2 = |\sfac_+|^2(k) +|\sfac_-(k)|^2\\
             &= -{\alpha\over \pi}\bigg({q_1\over kq_1}-{q_2\over kq_2} \bigg)^2.
\end{split}
\end{equation}
What is important for our discussion is that the IR-finite building blocks 
\begin{equation}
\begin{split}
  &\bbeta_0= \sum_\lambda |\Meu_\lambda|^2,\\
  &\bbeta_1(k)=  \sum\limits_{\lambda\sigma} |\Meu^{\rm 1-phot}_{\lambda\sigma}|^2 
                  -\sum\limits_{\sigma}  |\sfac_\sigma(k)|^2 
                   \sum\limits_{\lambda} |\Meu^{\rm Born}_\lambda|^2
\end{split}
\end{equation}
in the multiphoton distributions are all
in terms of $\sum\limits_{spin} |...|^2 $!!
We denoted: $\lambda$ = fermion helicities and $\sigma$ = photon helicity.

The above is to be contrasted with the analogous \Order{\alpha^1} case of CEEX
\begin{equation}
\begin{split}
\sigma &= \sum\limits_{n=0}^\infty\;
          \int\limits_{{ m_\gamma}} d\Phi_{n+2}\\
       &  \sum\limits_{\lambda,\sigma_1,...,\sigma_n}
          |e^{B({m_\gamma})}
          \Meu^{\lambda}_{n,\sigma_1,...,\sigma_n}(k_1,...,k_n)|^2,
\end{split}
\end{equation}
where the differential distributions
for $n_\gamma=0,1,2$ photons are the following:
\begin{equation}
\begin{split}
   &\Meu_{0}^{\lambda} = \hbeta_0^\lambda,\quad \lambda={\rm fermion helicities},\\
   &\Meu^\lambda_{1,\sigma_1}(k_1) 
              = \hbeta^\lambda_0 \sfac_{\sigma_1}(k_1)
               +\hbeta^\lambda_{1,\sigma_1}(k_1),\\
   &\Meu^\lambda_{2,\sigma_1,\sigma_2}(k_1,k_2) 
              = \hbeta^\lambda_0 \sfac_{\sigma_1}(k_1) \sfac_{\sigma_2}(k_2)\\
   &\qquad     +\hbeta^\lambda_{1,\sigma_1}(k_1)\sfac_{\sigma_2}(k_2) 
               +\hbeta^\lambda_{1,\sigma_2}(k_2)\sfac_{\sigma_1}(k_1)
\end{split}
\end{equation}
and the IR-finite building blocks are
\begin{equation}
\begin{split}
   &\hbeta^\lambda_0 = \big(e^{-B} \Meu^{\rm Born+Virt.}_{\lambda}\big)\big|_{{\cal O}(\alpha^1)},\\
   &\hbeta^{\lambda}_{1,\sigma}(k)=\Meu^\lambda_{1,\sigma}(k) - \hbeta^\lambda_0 \sfac_{\sigma}(k).
\end{split}
\end{equation}
As shown explicitly, this time
everything is in terms of $\Meu$-spin-amplitudes!
This is the basic difference between EEX/YFS and CEEX.
The complete expressions for
spin amplitudes with CEEX exponentiation, for any number of photons,
are shown in ref.~\cite{ceex1:1999} for the \Order{\alpha^1} case
and in ref.~\cite{ceex2:2000} for the \Order{\alpha^2} case.

%%%%%%%%%%%%%%%%%%%%%%%%%%%%%%%%%%%%%%%%%%%%%%%%%%%%%%%%%%%%%%%%%%%%%%%%%%%%%
\begin{figure}[!ht]
\centering
%---------------------------
\setlength{\unitlength}{0.1mm}
\begin{picture}(800,800)
%#####\put(0,0){\framebox( 800,800){ }}
%#####
\put( 0, 0){\makebox(0,0)[lb]{\epsfig{file=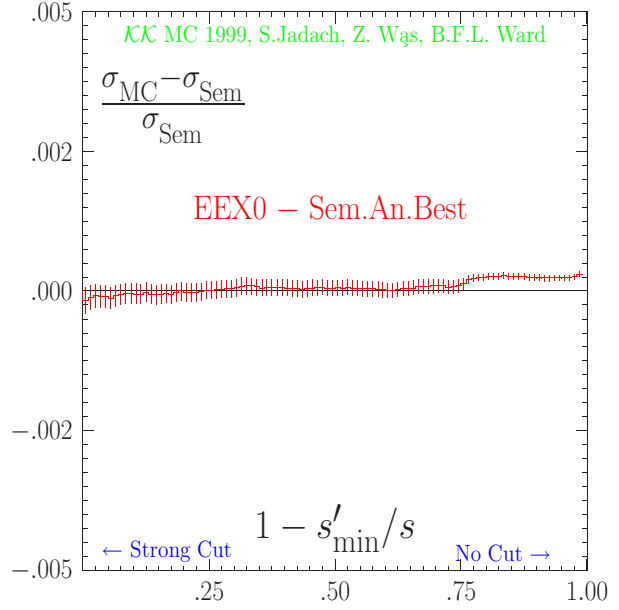,width=80mm,height=80mm}}}
\end{picture}
\vspace{-6mm}
\caption{
Baseline test of the technical precision.
}
\label{fig:technpr}
\end{figure}
%%%%%%%%%%%%%%%%%%%%%%%%%%%%%%%%%%%%%%%%%%%%%%%%%%%%%%%%%%%%%%%%%%%%%%%%%%%%%

%%%%%%%%%%%%%%%%%%%%%%%%%%%%%%%%%%%%%%%%%%%%%%%%%%%%%%%%%%%%%%%%%
\section{Monte Carlo numerical results}
%%%%%%%%%%%%%%%%%%%%%%%%%%%%%%%%%%%%%%%%%%%%%%%%%%%%%%%%%%%%%%%%%
The \Order{\alpha^2} CEEX-style matrix element is implemented in \KK MC
which simulates production of muon, tau and quark pairs.
Electrons (Bhabha scattering) and neutrino channels are not available.
The program includes for the optional use the older, EEX-style matrix element.
It is functionally similar to KORALZ~\cite{koralz4:1994} 
and the older KORALB~\cite{koralb2:1995} programs.
In Table~\ref{tab:compare} we provide the complete comparison of features of \KK MC
and the older programs.

%%%%%%%%%%%%%%%%%%%%%%%%%%%%%%%%%%%%%%%%%%%%%%%%%%%%%%%%%%%%%%%%%%%%%%%%%%%%%
\begin{figure*}[!ht]
\centering
%---------------------------
\setlength{\unitlength}{0.1mm}
\begin{picture}(1600,800)
%#####\put(0,0){\framebox( 1600,800){ }}
\put( 0, 0){\makebox(0,0)[lb]{\epsfig{file=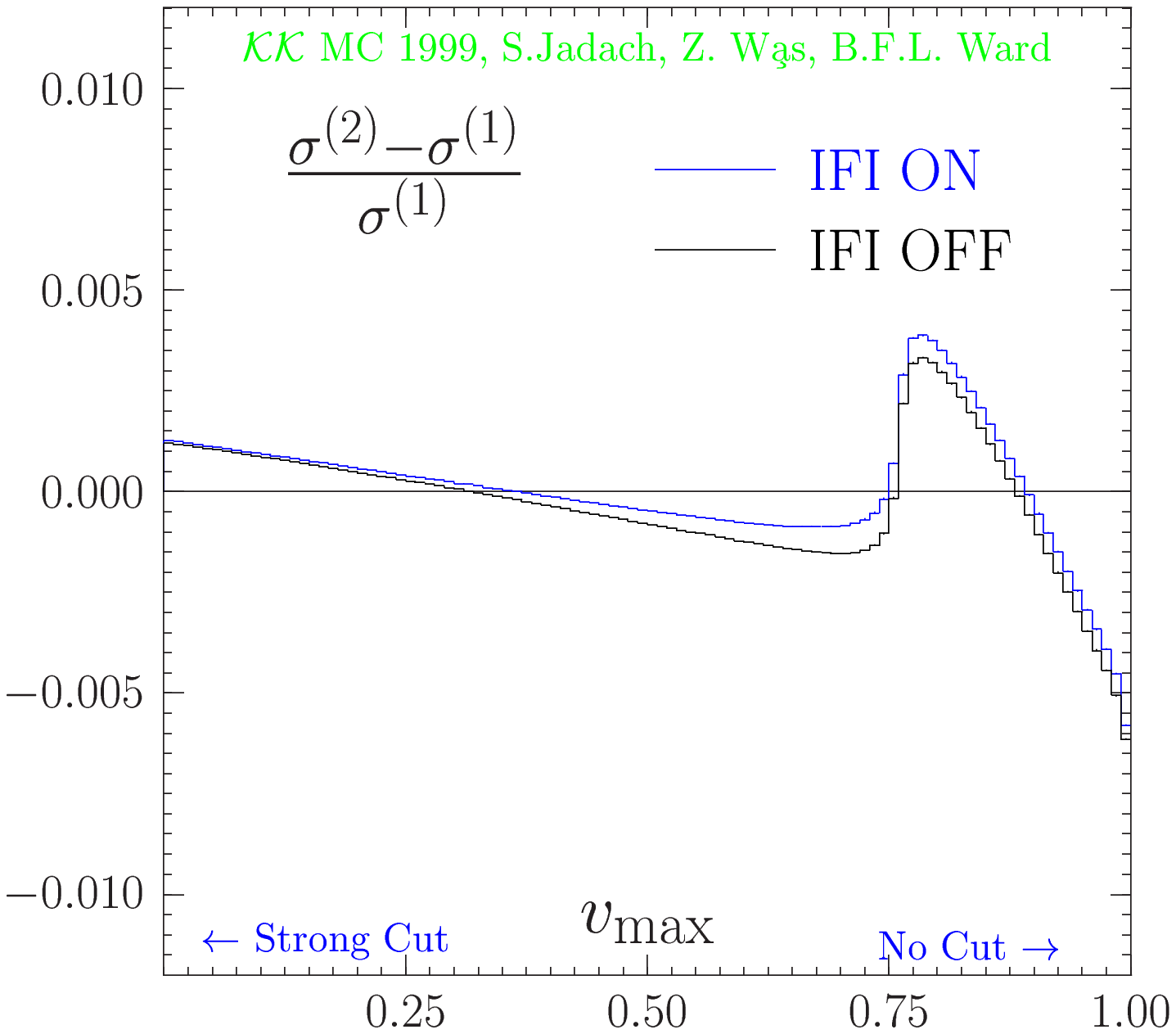,width=80mm,height=80mm}}}
%#####
\put( 800, 0){\makebox(0,0)[lb]{\epsfig{file=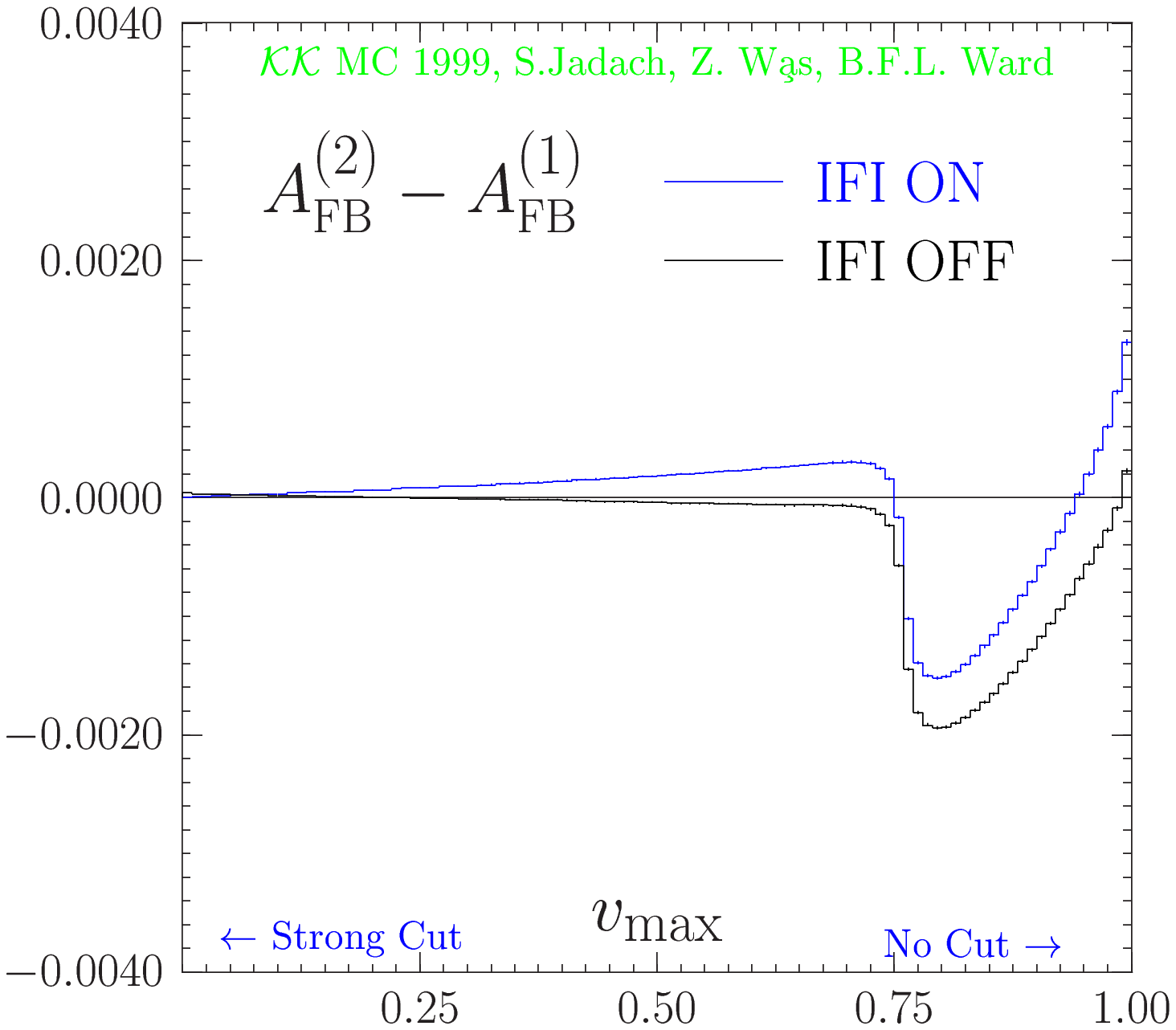,width=80mm,height=81mm}}}
\end{picture}
\vspace{-8mm}
\caption{ Test of the technical precision of \KK MC.
}
\label{fig:physpr}
\end{figure*}
%%%%%%%%%%%%%%%%%%%%%%%%%%%%%%%%%%%%%%%%%%%%%%%%%%%%%%%%%%%%%%%%%%%%%%%%%%%%%

%%%%%%%%%%%%%%%%%%%%%%%%%%%%%%%%%%%%%%%%%%%%%%%%%%%%%%%%%%%%%%%%%
\subsection{Technical precision}
For the new MC program of the high complexity like \KK MC it is important
to check very precisely the overall normalization.
This is the cornerstone of the evaluation of the {\em technical precision}
of the program, especially for  \KK MC which is aimed at the end of testing
at the total precision of 0.1\%.
In Fig.~\ref{fig:technpr} we present the comparison of the \KK MC with simple
semianalytical integration for the total cross section, as a function of
the minimum mass $\sqrt{s'_{\min}}$ of the final muon pair.
It is done for muon-pair final state at $\sqrt{s}=200GeV$.
For $\sqrt{s'_{\min}}\to \sqrt{s}$, when emission of hard photons is suppressed,
there is an agreement $<0.02\%$ between \KK MC and the analytical calculation.
For $\sqrt{s'_{\min}}<M_Z$ the on-shell Z-boson production
due to emission of the hard initial state radiation (ISR),
the so called Z radiative return (ZRR), is allowed kinematically.
Even in this case (more sensible to higher orders) the agreement $<0.02\%$ is reached.
For the above exercise we used the simplified \Order{\alpha^0} CEEX matrix element,
because in this case the precise phase analytical integration is relatively easy.
%%%%%%%%%%%%%%%%%%%%%%%%%%%%%%%%%%%%%%%%%%%%%%%%%%%%%%%%%%%%%%%%%
\subsection{Physical precision}
The equally important component of the overall error is the physical error
which we estimate conservatively as the half of the difference
\Order{\alpha^2}$-$\Order{\alpha^1}.
In Fig.~\ref{fig:physpr} we show  the corresponding result for the total
cross section and charge asymmetry for $\sqrt{s}=189GeV$ as a function
of the cut on energies of all photons 
($s'_{\min}>s$ limits the total photon energy.)
We obtain in this way the estimate $0.2\%$ for the physical precision 
of the total cross section and $0.1\%$ for the charge asymmetry.
Both plots in Fig.~\ref{fig:physpr} show as expected
strong variation at the position of the ZRR.
This precision is good enough for the LEP2 combined data.

%%%%%%%%%%%%%%%%%%%%%%%%%%%%%%%%%%%%%%%%%%%%%%%%%%%%%%%%%%%%%%%%%%%%%%%%%%%%%
\begin{figure*}[!ht]
\centering
%---------------------------
\setlength{\unitlength}{0.1mm}
\begin{picture}(1600,800)
%#####\put(0,0){\framebox( 1600,800){ }}
\put( 0, 0){\makebox(0,0)[lb]{\epsfig{file=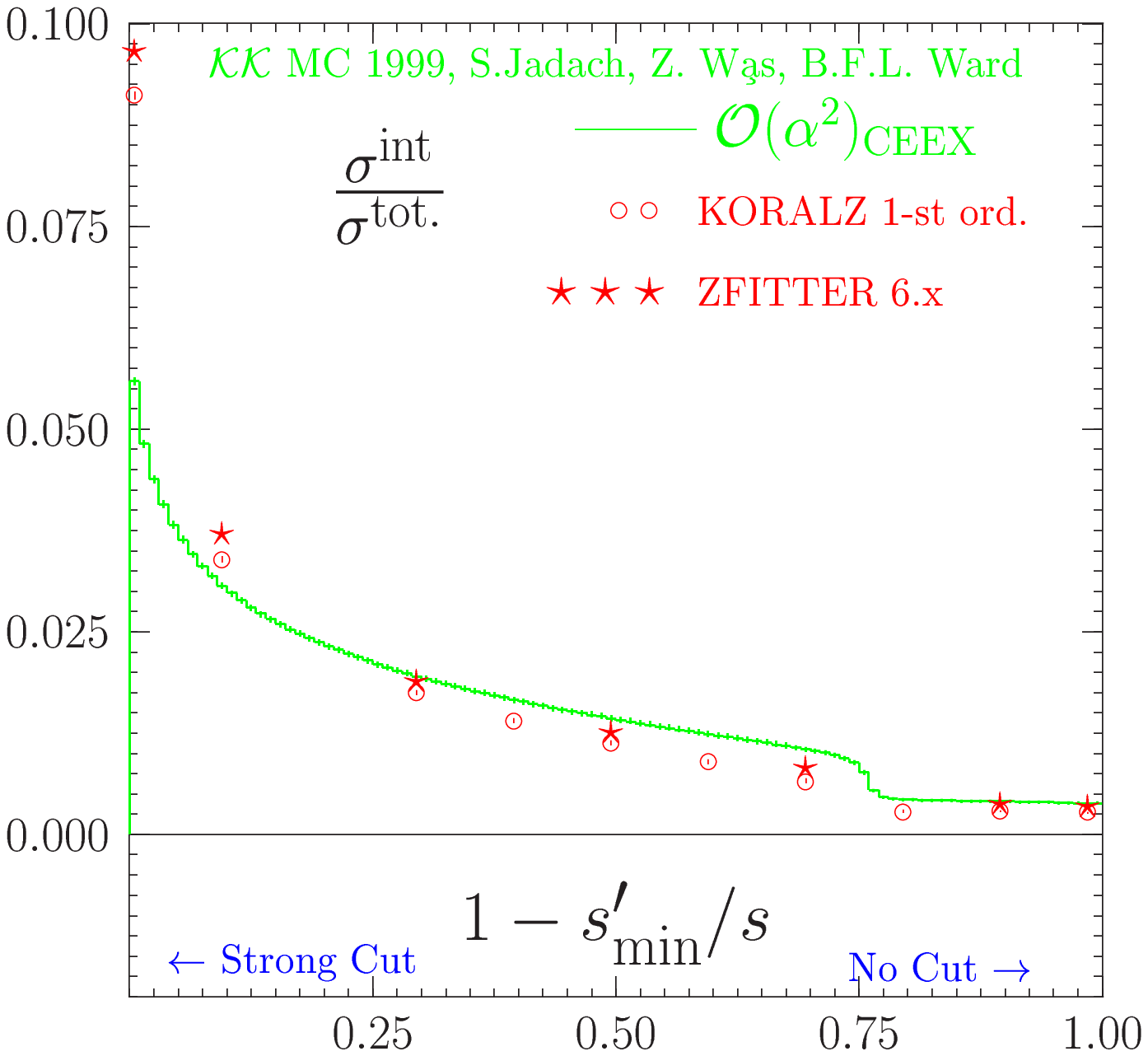,width=80mm,height=80mm}}}
%#####
\put( 800, 0){\makebox(0,0)[lb]{\epsfig{file=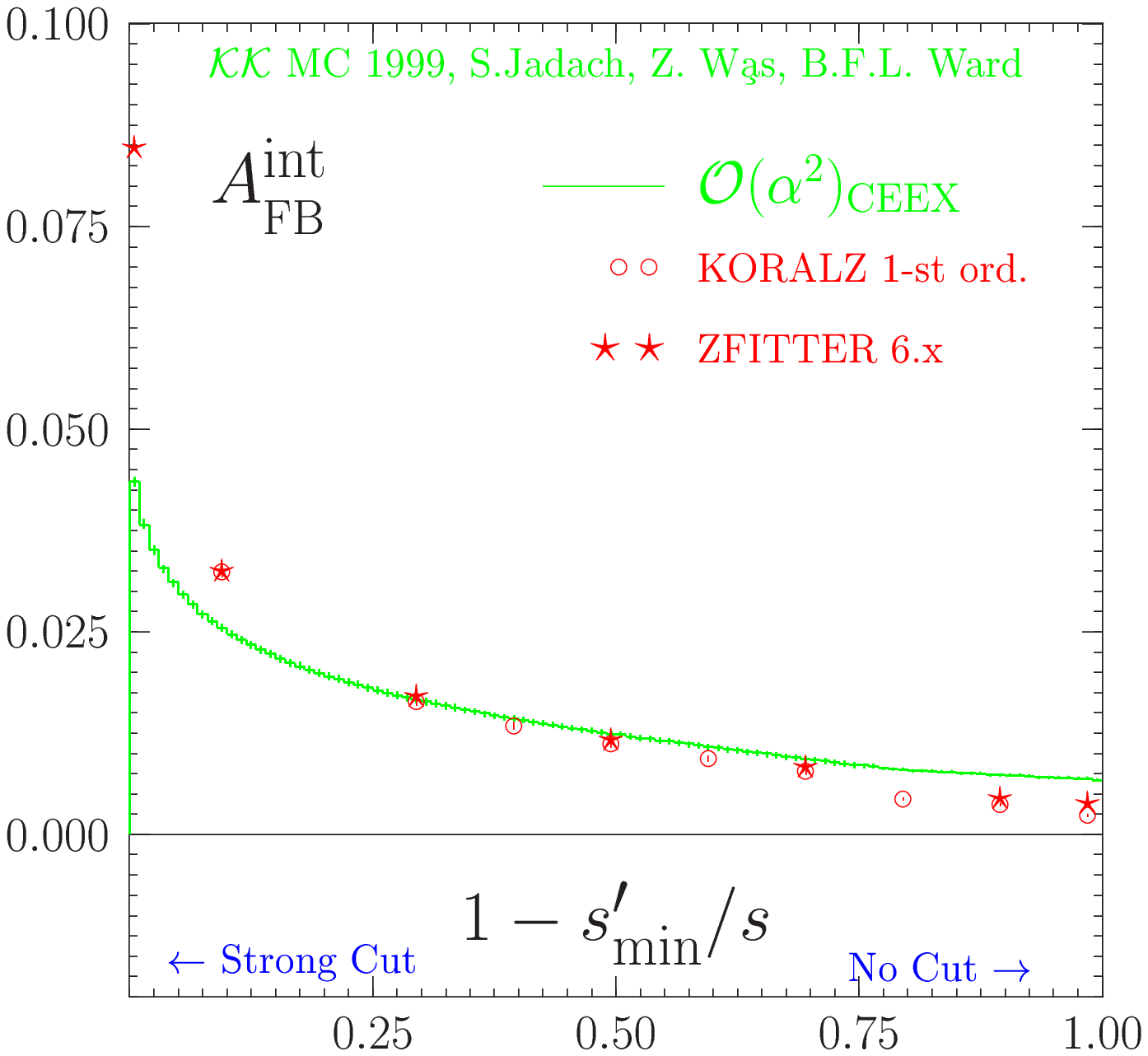,width=80mm,height=81mm}}}
\end{picture}
\vspace{-8mm}
\caption{
 The effect of the initial-final state QED interference in total cross-section
 and charge asymmetry.
}
\label{fig:IFI}
\end{figure*}
%%%%%%%%%%%%%%%%%%%%%%%%%%%%%%%%%%%%%%%%%%%%%%%%%%%%%%%%%%%%%%%%%%%%%%%%%%%%%

%%%%%%%%%%%%%%%%%%%%%%%%%%%%%%%%%%%%%%%%%%%%%%%%%%%%%%%%%%%%%%%%%
\subsection{Initial-final state QED interference}
One important benefit from CEEX with respect to the older EEX
is the inclusion of the Initial-Final state QED Interference (IFI).
The effect of IFI is comparable with the precision of the LEP2 combined data
and should be under good control.
Results of our analysis of the size of IFI at LEP2 energies ($\sqrt{s}=189GeV$)
are shown in Fig.~\ref{fig:IFI}.
In this figure we compare the CEEX result of \KK MC first of
all with the result of KORALZ which is
run in the \Order{\alpha^1} mode without exponentiation
(IFI is neglected for KORALZ with the EEX matrix element.)
The \Order{\alpha^1} IFI contribution from  KORALZ was extensively cross-checked
in the past with the dedicated semi-analytical calculations~\cite{afb-prd:1991},
it is therefore a good reference and starting point.
As we see the IFI contribution of CEEX differs slightly from the pure \Order{\alpha^1}
result. 
It is related to exponentiation which makes the angular dependence (in the muon scattering angle)
of the IFI contribution less sharp and it is also due to convolution of IFI
with the \Order{\alpha^2} ISR.
The expected modification of the interference correction due to higher
orders is about 20\% for cross section and asymmetry, if ZRR is excluded,
(the size of ISR correction in the  cross section)
and it is indeed of this size.
Apparently, this principle works also in the case of ZRR included, 
remembering that in this case ISR correction is 100\% or more.
However, we feel that this case requires further study.
%%%The expected modification of the cross section is about 20\%
%%%(the size of ISR) and for the total cross section it is of this size.
%%%For the charge asymmetry the effect of exponentiation and ISR seems
%%%to be bigger than expected for the ZRR.
%%%This requires further study.
We have also included results of the semianalytical program
ZFITTER~\cite{zfitter6:1999} in our plots%
\footnote{ We would like to thank D. Bardin for providing us results from ZFITTER.}.
They agree well with the \Order{\alpha^1} IFI of KORALZ.
This is expected because they are without exponentiation.

%%%%%%%%%%%%%%%%%%%%%%%%%%%%%%%%%%%%%%%%%%%%%%%%%%%%%%%%%%%%%%%%%%%%%%%%%%%%%%%%%%%%%%%%%%%%%
%%%%%%%%%%%%%%%%%%%%%%%%%%%%%%%%%%%%%%%%%%%%%%%%%%%%%%%%%%%%%%%%%%%%%%%%%%%%%%%%%%%%%%%%%%%%%
\section{Outlook and summary}
%%%%%%%%%%%%%%%%%%%%%%%%%%%%%%%%%%%%%%%%%%%%%%%%%%%%%%%%%%%%%%%%%%%%%%%%%%%%%%%%%%%%%%%%%%%%%
The most important new features in the present CEEX are the
ISR-FSR interference, the second-order subleading corrections, and the exact matrix
element for two hard photons.
This makes CEEX already a unique source of SM predictions for the LEP2 physics program
and for the LC physics program.
Note that for these the electroweak correction library has to be reexamined at LC energies.
The most important omission in the present version is the lack of neutrino and electron
channels.
Let us stress that the present program is an excellent starting platform
for the construction of the second-order Bhabha MC generator based on CEEX exponentiation.
We hope to be able to include the Bhabha and neutrino channels soon, 
possibly in the next version.
The other important directions for the development are 
the inclusion of the exact matrix element for three hard photons, together
with virtual corrections up to \Order{\alpha^3L^3} and the
emission of the light fermion pairs.
The inclusion of the $W^+W^-$ and $t\bar{t}$ final states is still in a farther perspective.

%%%%%%%%%%%%%%%%%%%%%%%%%%%%%%%%%%%%%%%%%%%%%%%%%%%%%%%%%%%%%%%%%%%%%%%%%%%%%%%%%%%%%%%%%%%%%

%%%%%%%%%%%%%%%%%%%%%%%%%%%%%%%%%%%%%%%%%%%%%%%%%%%%%%%%%%%%%%%%%%%%%%%%%%%%
%%%%%%%%%%%%%%%%%%%%%%%%%%%%%%%%%%%%%%%%%%%%%%%%%%%%%%%%%%%%%%%%%%%%%%%%%%%%
%\bibliographystyle{prsty}
%\bibliographystyle{plain}
%\bibliography{KK}

\begin{thebibliography}{10}

\bibitem{ceex1:1999}
S. Jadach, B.~F.~L. Ward, and Z. W\c{a}s, Phys. Lett. {\bf B449},  97  (1999).

\bibitem{gps:1998}
S. Jadach, B.~F.~L. Ward, and Z. Was, Global positioning of spin GPS scheme for
  half spin massive spinors, 1998, preprint hep-ph/9905452, CERN-TH-98-235,
  submitted to Eur. J. Phys. C.

\bibitem{ceex2:2000}
S. Jadach, B.~F.~L. Ward, and Z. W\c{a}s, Coherent Exclusive Exponentiation For
  Precision Monte Carlo Calculations, 2000, preprint
  CERN-TH/2000-087,UTHEP-99-09-01.

\bibitem{kkcpc:1999}
S. Jadach, Z. W\c{a}s, and B.~F.~L. Ward, The Precision Monte Carlo Event
  Generator ${\cal KK}$ For Two-Fermion Final States In $e^+e^-$ Collisions,
  Computer Physics Communications in print, 2000, preprint DESY-99-106,
  CERN-TH/99-235, UTHEP-99-08-01, source version 4.13 available from
  http://home.cern.ch/jadach/.

\bibitem{sussex:1989}
S. Jadach and B. Ward,  in {\em Electroweak Physics}, edited by N. Dombey and
  F. Boudjema (Plenum Publ. Co., London, 1989), \uppercase{P}roc. of Sussex
  University Conference.

\bibitem{yfs:1961}
D.~R. Yennie, S. Frautschi, and H. Suura, Ann. Phys. (NY) {\bf 13},  379
  (1961).

\bibitem{yfs2:1990}
S. Jadach and B.~F.~L. Ward, Comput. Phys. Commun. {\bf 56},  351  (1990).

\bibitem{koralz4:1994}
S. Jadach, B.~F.~L. Ward, and Z. W\c{a}s, Comput. Phys. Commun. {\bf 79},  503
  (1994).

\bibitem{koralw:1998}
S. Jadach {\it et~al.}, Comput. Phys. Commun. {\bf 119},  272  (1999).

\bibitem{yfsww:1996}
S. Jadach, W. P{\l}aczek, M. Skrzypek, and B.~F.~L. Ward, Phys. Rev. {\bf D54},
   5434  (1996).

\bibitem{bhlumi4:1996}
S. Jadach {\it et~al.}, Comput. Phys. Commun. {\bf 102},  229  (1997).

\bibitem{bhwide:1997}
S. Jadach, W. P\l{}aczek, and B.~F.~L. Ward, Phys. Lett. {\bf B390},  298
  (1997), also hep-ph/9608412; The Monte Carlo program BHWIDE is available from
  {\tt http://hephp01.phys.utk.edu/pub/BHWIDE}.

\bibitem{koralb2:1995}
S. Jadach and Z. Was, Comput. Phys. Commun. {\bf 85},  453  (1995).

\bibitem{afb-prd:1991}
S.Jadach and Z.W\c{a}s, Phys. Rev. {\bf D41},  1425  (1990).

\bibitem{zfitter6:1999}
D. Bardin {\it et~al.}, ZFITTER v.6.21: A Semianalytical program for fermion
  pair production in e+ e- annihilation, 1999, e-print: hep-ph/9908433.

\end{thebibliography}
%%%%%%%%%%%%%%%%%%%%%%%%%%%%%%%%%%%%%%%%%%%%%%%%%%%%%%%%%%%%%%%%%%%%%%%%%%%%
%%%%%%%%%%%%%%%%%%%%%%%%%%%%%%%%%%%%%%%%%%%%%%%%%%%%%%%%%%%%%%%%%%%%%%%%%%%%

\end{document}